\documentclass[aps,prd,twocolumn,superscriptaddress]{revtex4-2}

\usepackage{graphicx}
\usepackage{color}
\usepackage[dvipsnames]{xcolor}
\usepackage{rotating}
\usepackage{bigints}
\usepackage{enumerate}
\usepackage{lipsum}

\usepackage[colorlinks=true,backref=false, linktocpage=true,
citecolor=blue,urlcolor=blue,linkcolor=blue,pdfpagemode=UseOutlines,pdfstartview=FitH,bookmarksnumbered=true,bookmarksopen]{hyperref}

\newcommand{\beq}{\begin{equation}}
\newcommand{\eeq}{\end{equation}}
\newcommand{\bea}{\begin{eqnarray}}
\newcommand{\eea}{\end{eqnarray}}
\newcommand \hmu {\hat{\mu}}

\newcommand{\calO}{{\cal O}}

\newcommand{\MeV}{\, {\rm MeV}}
\newcommand{\lat}[2]{{{#1}^3\!\!\times\!\!{#2}}}

\newenvironment{paolo}{\bf \color{MidnightBlue}}{}
\newcommand{\bepa}{\begin{paolo}}
\newcommand{\enpa}{\end{paolo}}

\begin{document}


\title{Lattice QCD equation of state at finite chemical potential from an alternative 
expansion scheme}



\author{S. Bors\'anyi}
\affiliation{University of Wuppertal, Department of Physics, Wuppertal D-42119, Germany}

\author{Z. Fodor}
\affiliation{Pennsylvania State University, Department of Physics, State College, PA 16801, USA}
\affiliation{University of Wuppertal, Department of Physics, Wuppertal D-42119, Germany}
\affiliation{Inst.  for Theoretical Physics, ELTE E\"otv\"os Lor\' and University, P\'azm\'any P. s\'et\'any 1/A, H-1117 Budapest, Hungary}
\affiliation{J\"ulich Supercomputing Centre, Forschungszentrum J\"ulich, D-52425 J\"ulich, Germany}

\author{J. N. Guenther}
\affiliation{Aix Marseille Univ., Universit\'e de Toulon, CNRS,
CPT, Marseille, France}

\author{R. Kara}
\affiliation{University of Wuppertal, Department of Physics, Wuppertal D-42119, Germany}

\author{S. D. Katz}
\affiliation{E{\"o}tv{\"o}s University, Budapest 1117, Hungary} 

\author{P. Parotto}
\email[Corresponding author: ]{parotto@uni-wuppertal.de}
\affiliation{University of Wuppertal, Department of Physics, Wuppertal D-42119, Germany}

\author{A. P\'asztor}
\affiliation{E{\"o}tv{\"o}s University, Budapest 1117, Hungary}

\author{C. Ratti}
\affiliation{Department of Physics, University of Houston, Houston, TX 77204, USA}

\author{K. K. Szab\'o}
\affiliation{University of Wuppertal, Department of Physics, Wuppertal D-42119, Germany}
\affiliation{J\"ulich Supercomputing Centre, Forschungszentrum J\"ulich, D-52425 J\"ulich, Germany}


\date{\today}

\begin{abstract}

Taylor expansion of the equation of state of QCD 
suffers from shortcomings at chemical potentials $\mu_B \geq (2-2.5)T$.
First, one faces difficulties inherent in performing such an expansion 
with a limited number of coefficients; second,
higher order coefficients determined from 
lattice calculations suffer from a poor signal-to-noise ratio.
In this work, we present a novel scheme for extrapolating 
the equation of state of QCD to finite, real chemical 
potential that can extend its reach further than previous methods.
We present continuum extrapolated lattice results for the new
expansion coefficients and show the thermodynamic observables up
to $\mu_B/T\le3.5$.

\end{abstract}


\maketitle

\section{Introduction}

The phase diagram of Quantum Chromodynamics 
(QCD) is an open field of investigation, which is 
currently at the center of intense efforts from the 
theoretical and experimental communities alike. 
At vanishing baryon density, the transition between
confined and deconfined matter is known to be an
analytic crossover \cite{Aoki:2006we}. This knowledge was
gained through lattice simulations, which represent a systematically
improvable method to solve equilibrium QCD. 
Although at finite baryon density lattice QCD faces a sign
problem, numerous results have been published
for moderate chemical potentials in recent years
\cite{Guenther:2020vqg,Ratti:2018ksb}.  New techniques that allow for direct
simulations at finite chemical potential in the presence of a
sign problem are the subject of intense investigation. Promising
results are available from Lefschetz thimbles \cite{Cristoforetti:2012su,Alexandru:2015xva,Fukuma:2019uot},
the Complex Langevin equation \cite{Aarts:2012ft,Sexty:2013ica,Scherzer:2019lrh}
or reweighting-based methods \cite{Giordano:2020roi}.
However, these novel approaches cannot be applied to large
scale QCD simulations with physical quark masses yet.

There are many indirect lattice methods to study QCD
at finite density. These are based on the known analytic
feature of the QCD free energy at zero baryo-chemical potential ($\mu_B$).
The conceptually simplest method is the Taylor expansion, where
the leading $\mu_B$-derivatives of the relevant observables
are calculated
\cite{Allton:2002zi,Allton:2005gk,Borsanyi:2012cr,Bazavov:2017dus}.
These derivatives were often calculated for chiral observables,
and the $\mu_B$ dependence of the transition temperature could
be extracted
\cite{Kaczmarek:2011zz,Endrodi:2011gv,Bonati:2018nut,Bazavov:2018mes}.

An additional twist to the Taylor method is the observation that these
coefficients can be efficiently calculated by simulating at
imaginary values of the chemical potential(s), besides $\mu_B=0$
\cite{DElia:2016jqh,Borsanyi:2018grb}. The use of imaginary chemical
potentials is motivated by the analytic crossover at $\mu_B=0$, from
which the smooth behavior of the thermodynamic observables 
as a function of $\mu_B^2$ follows. This approach has been popular
for more than a decade \cite{deForcrand:2002hgr,DElia:2002tig,DElia:2007bkz,Cea:2009ba,Cea:2010bz,Bonati:2013tqa,Bonati:2018nut}.
This possibility opened an avenue towards finite density physics, which is
often referred to as analytical continuation.
This name suggests that, besides the computation of the Taylor expansion
coefficients, other extrapolation schemes can be established.  The success of
this method was most visible in the study of the QCD transition line, where 
continuum extrapolated results are available for the leading $\mu_B$ dependence
\cite{Bonati:2015bha,Cea:2015cya,Bellwied:2015rza,Bonati:2018nut}, and recently
also for the
next-to-leading coefficient \cite{Borsanyi:2020fev}.
On the other hand, the Taylor series is just one of the possible
expansion or continuation schemes.
For example, the Pad\'e summation was also considered in
the context of QCD thermodynamics
\cite{Karsch:2010hm,Cea:2012ev,Datta:2016ukp,Pasztor:2020dur,Schmidt:2021pey}
.

The knowledge of the features of the QCD phase 
diagram from lattice simulations is currently limited
to small chemical potentials and data are mostly available
only in the transition region. We have to mention, though, that
at higher temperatures resummed perturbation theory has
provided a quantitative description of the chemical
potential dependence of several observables \cite{Mogliacci:2013mca,Haque:2013sja,Haque:2020eyj}.
Dedicated lattice studies have bridged the gap between
the transition region and perturbative temperatures and
found perfect agreement \cite{Bellwied:2015lba,Ding:2015fca}.

\begin{figure*}[t]
  \makebox[\textwidth][c]
{
\includegraphics[width=0.39\textwidth]{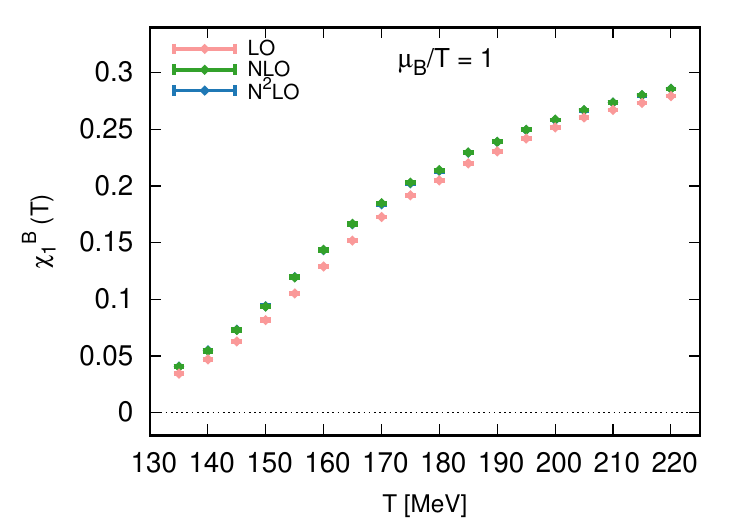} \hspace{-10mm}
\includegraphics[width=0.39\textwidth]{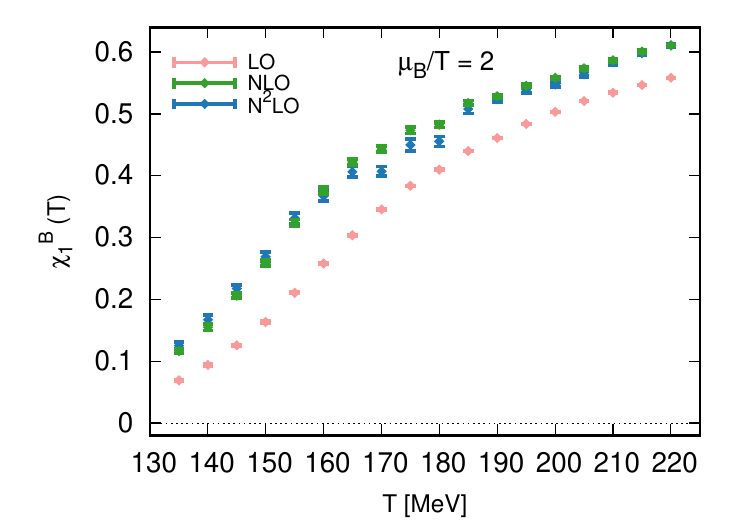} \hspace{-10mm}
\includegraphics[width=0.39\textwidth]{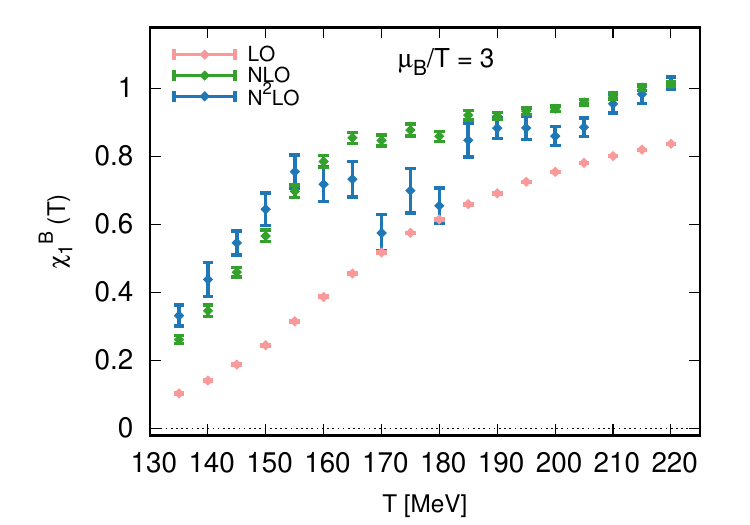}}%
\caption{Baryon density from a Taylor expansion with 
the coefficients in Ref.~\cite{Borsanyi:2018grb}, at 
$\mu_B/T = 1, 2, 3$, as a function of the temperature. Different colors correspond to the 
order to which the expansion is carried out.}
\label{fig:chiB1Taylor}
\end{figure*}

On the experimental side, several heavy-ion collision 
programs are in place, with the explicit intent of 
mapping out the phase structure of strongly 
interacting matter. An important tool for the 
theoretical interpretation of the results from
these experiments are hydrodynamic 
simulations, which describe the evolution of the system created in the collisions. These simulations need the equation of state of QCD as an input, in the whole range of temperatures and densities covered in the experiments. Recently, a Bayesian analysis based on a systematic comparison between heavy-ion data and theoretical predictions showed that the posterior distribution over possible equations of states is compatible with the one calculated on the lattice \cite{Pratt:2015zsa}. For this reason, the equation of state at finite density is a crucial ingredient to understand the nature of strongly interacting matter, and to support the experimental program.

The equation of state at vanishing chemical potential 
has been known now for several years over a broad range of temperatures
\cite{Borsanyi:2010cj,Borsanyi:2013bia,Bazavov:2014pvz}.
The first continuum extrapolated extension to finite $\mu_B$ using
the Taylor method in Ref.~\cite{Borsanyi:2012cr} was followed by
several works with the intent of extending these results to higher $\mu_B$ by adding more terms in the Taylor series
\cite{Bazavov:2017dus,Gunther:2016vcp}. Currently, even the eighth
$\mu_B$-derivative of the QCD pressure is available with modest precision
from lattice simulations \cite{Borsanyi:2018grb,Bazavov:2020bjn}.  Very
recently, similar results were found by solving a QCD-assisted effective theory
with functional methods \cite{Fu:2021oaw}.

In this work, we propose a new extrapolation scheme to finite density QCD.
We intend to remedy some shortcomings of the Taylor-based
equation of state, e.g. the extrapolation through a crossover boundary.
We discuss this issue and suggest a solution in Section \ref{sec:motivation}.
The formalism for our method is worked out in Section
\ref{sec:formalism}. We then compute the new observables from
lattice simulations in Section \ref{sec:lattice} and perform their continuum
estimate. Using this lattice input, we construct the finite density
thermodynamic functions in Section \ref{sec:thermo}. We conclude and
discuss further aspects in Section \ref{sec:conclusions}.

\section{Motivation\label{sec:motivation}}


The knowledge of the equation of state from lattice 
simulations commonly consists of the established
$\mu_B=0$ result \cite{Borsanyi:2013bia,Bazavov:2014pvz}
and the Taylor expansion coefficients of the pressure around 
$\mu_B=0$
\begin{equation}
\frac{p(T,\mu_B)}{T^4}
=
\sum_{n=0} \frac{1}{(2n)!} \chi^B_{2n}(T,0) \left(\frac{\mu_B}{T}\right)^{2n},
\label{eq:ptaylor}
\end{equation}
where $\chi^B_j$ are the $j$-th derivatives of the normalized pressure:
\begin{equation}
\chi^{B}_{j}(T,\mu_B) = 
\left(\frac{\partial}{\partial \mu_B/T}\right)^j
 \frac{p(T,\mu_B)}{T^4} \, \, .
\end{equation}
Besides diagonal coefficients, one can also define off-diagonal correlators between different conserved charges in QCD. Correlators between baryon number and strangeness, which we will need in our procedure, are defined as follows
\begin{equation}
\chi^{BS}_{jk}(T,\mu_B) = 
\left(\frac{\partial}{\partial \mu_B/T}\right)^j
\left(\frac{\partial}{\partial \mu_S/T}\right)^k
 \frac{p(T,\mu_B)}{T^4} \, \, .
\end{equation}
Such correlators have phenomenological relevance \cite{Bellwied:2019pxh} and they can also be used to extrapolate the equation of state of QCD in the full, four-dimensional phase diagram at finite $T,~\mu_B,~\mu_S,~\mu_Q$ \cite{Noronha-Hostler:2019ayj,Monnai:2019hkn}.
We will use the $\hmu_i=\mu_i/T$ shorthand notation in this manuscript.

Currently, results for the expansion coefficients
are available up to order $\calO (\mu_B^6)$
\cite{Bazavov:2020bjn,Borsanyi:2018grb}. The region 
of validity of the resulting expansion is usually 
determined by the range in chemical potential within 
which an apparent convergence is achieved with the
available coefficients. This is stated to be 
$\mu_B/T \lesssim 2-2.5$ \cite{Gunther:2016vcp,Bazavov:2017dus}.

High order derivatives of the pressure are notoriously 
difficult to calculate, as they suffer from a low signal-to-noise ratio. This is because their direct determination
involves large cancellations of different terms 
containing derivatives of the Dirac operator \cite{Bellwied:2015lba}.
Moreover, studies of chiral models revealed that the structure of the
temperature dependence of such observables becomes more and 
more complex when higher orders are considered \cite{Friman:2011pf}.
It was pointed out in Ref.~\cite{Borsanyi:2018grb} that 
the linear $\mu_B^2$-dependence of the crossover temperature may
explain the basic structure.

This may explain why including one more term in a truncated Taylor series will
not always improve the convergence.  On the contrary, pathological behavior --
namely non-monotonicity in the $T$- or $\mu_B$-dependence -- appears in the
extrapolated thermodynamic quantities at chemical potentials beyond $\mu_B/T
\lesssim 2-2.5$. This is due to the fact that, for large enough $\mu_B/T$, the
observables at finite chemical potential are dictated by the $\mu_B=0$
temperature dependence of the last coefficient included in the expansion.
Hence, the structures appearing around the QCD transition temperature in higher
order coefficients are ``translated'' into the finite-$\mu_B$ behavior of e.g.,
the entropy, baryon density, etc.

Another inherent problem with the Taylor expansion is the fact that it is carried
out at constant temperature. This means that the values of the coefficients
at $\mu_B=0$ and a certain temperature $T$, 
determine the equation of state at the same $T$ at finite $\mu_B$, while the pseudo-critical
temperature $T_{pc}$ might have varied considerably. While a sufficiently
large number of expansion coefficients would lead to smooth extrapolated functions, even though the Taylor coefficients themselves present a complex structure around the transition temperature, the problem here is rather practical. A scheme that could work
with fewer coefficients would be much preferable from the numerical cost point of view.

\begin{figure}[h]
\includegraphics[width=\linewidth]{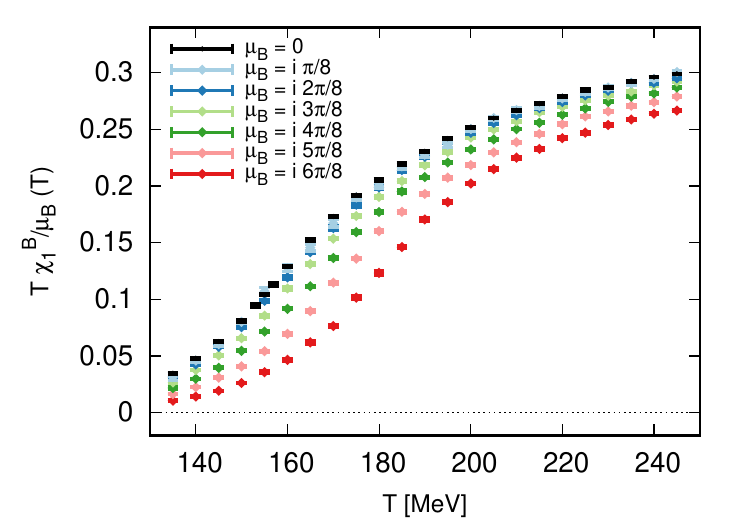}
\includegraphics[width=\linewidth]{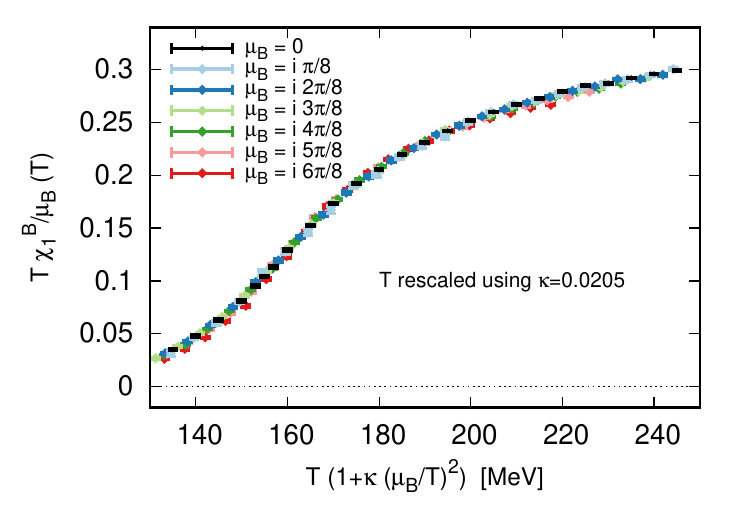}
\caption{Upper panel: The (imaginary) baryon density at simulated 
(imaginary) baryon chemical potentials, divided by the 
chemical potential. The points at $\mu_B=0$ (black)
show the second baryon susceptibility $\chi_2^B(T)$. Lower panel: same
curves as in the upper panel, with a temperature rescaled in accordance to
Eq.~\eqref{eq:Tprime1} with $\kappa=0.0205$.}
\label{fig:shift_B2}
\end{figure}

\begin{figure}[h]
\includegraphics[width=\linewidth]{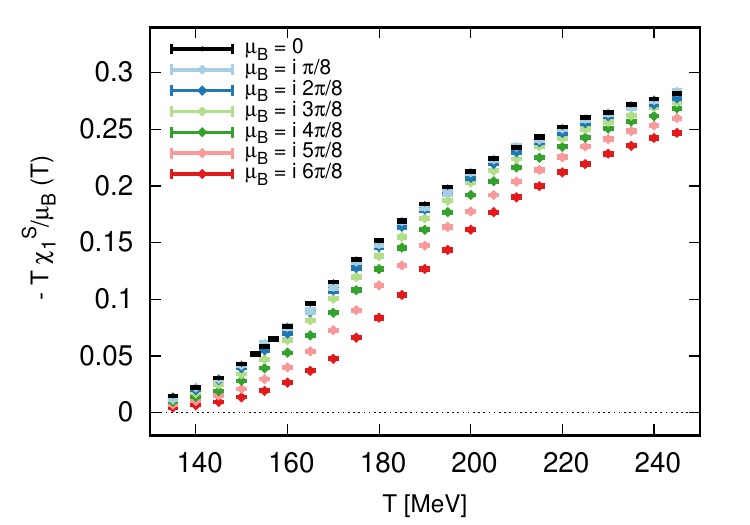}
\includegraphics[width=\linewidth]{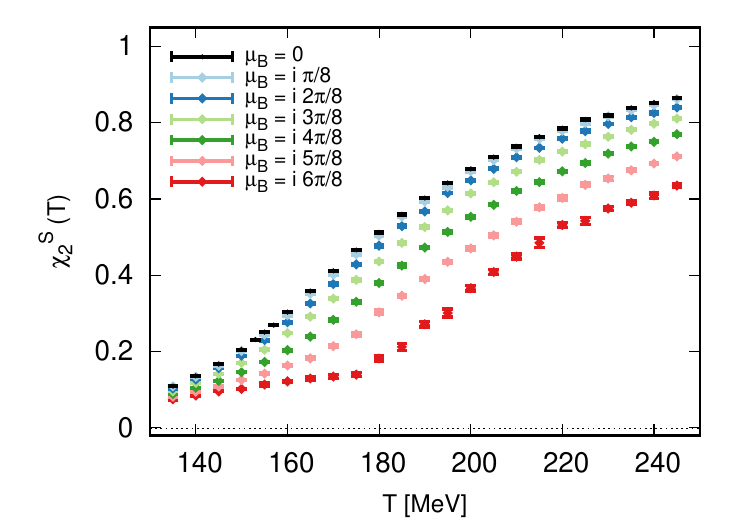}
\caption{The (imaginary) strangeness density divided 
by the baryon chemical potential (upper panel) and the 
second strangeness susceptibility (lower panel) at 
simulated (imaginary) baryon chemical potentials. The 
points at $\mu_B=0$ (black) show the baryon-
strangeness correlator $\chi_{11}^{BS}(T)$ (upper 
panel) and the second strangeness susceptibility 
$\chi_2^S(T)$ (lower panel), respectively.}
\label{fig:shift_BS_S2}
\end{figure}

In Fig.~\ref{fig:chiB1Taylor} we show the baryon density $n_B(T)$ obtained from
a Taylor expansion with the coefficients in  Ref.~\cite{Borsanyi:2018grb}, at
$\hmu_B = 1, 2, 3$. The extrapolation is shown including an increasing number
of coefficients, to show the effect of higher-order ones. 
The leading-order (LO) and higher truncations
refer to $\sim \hmu_B \partial n_B(T)/\partial\hmu_B$, or
$\sim \frac16 \hmu_B^3 \partial^3 n_B(T)/\partial\hmu_B^3$, etc.  being
the last term in the expansion. The derivatives are taken at $\mu_B=0$.

While at $\hmu_B=1$ apparent convergence is achieved at the NLO level, for
higher chemical potential this is not the case.  Especially at $\hmu_B=3$, the
inclusion of all the coefficients in Ref.~\cite{Borsanyi:2018grb} causes
unphysical non-monotonic behavior. Ultimately, a pathological behaviour
could also come from a finite radius of convergence. Incidentally,
recent estimates on coarse lattices \cite{Giordano:2019slo,Giordano:2019gev},
but also universality arguments \cite{Connelly:2020pno} place the 
convergence in the same ball-park in $\mu_B$.

In this work, we present an alternative summation 
scheme which can better cope with the fact that the 
QCD transition temperature presents a $\mu_B$-dependence. 

We start from the observation that we made while working
with imaginary values of the chemical potentials in an earlier
work on analytical continuation. In the upper panel of Fig.~\ref{fig:shift_B2} 
we show temperature scans of the quantity 
$n_B(T)/\hmu_B = \chi^B_1(T,\hmu_B)/\hmu_B$ for several fixed imaginary 
$\mu_B/T$ ratios. The $0/0$ limit at $\mu_B=0$ can be easily resolved and 
equals $\chi_2^B(T)$.

The $T$-dependence of the normalized baryon density at 
finite chemical potential appears to be simply shifted/rescaled
towards higher temperatures from the $\mu_B=0$ 
results for $\chi_2^B$. This behavior is more apparent 
in the vicinity of the transition, where the slope of 
these curves is larger. At very large, as well as at very 
low temperatures a simple shift cannot describe the 
physics, since the curves become extremely flat. 
A simple rescaling of temperatures can be described as:
\begin{equation} \label{eq:shift1}
\frac{\chi_1^B(T,\hmu_B)}{\hmu_B} = \chi_2^B (T^\prime,0) \, \,  , 
\end{equation}
where the actual temperature difference can be expressed through a 
$\mu_B$-dependent rescaling factor that we write for simplicity as
\begin{equation} \label{eq:Tprime1}
T^\prime = T \left( 1 + \kappa \hmu_B^2 \right) \, \, .
\end{equation}
In the lower panel of Fig.~\ref{fig:shift_B2} we show a version of the 
curves in the upper panel, where all the finite-$\hat{\mu}_B$ curves
have been shifted in accordance to Eq.~\eqref{eq:Tprime1} with
$\kappa=0.0205$. Remarkably, we note how well the curves are 
superimposed to each other, even with the simple assumption of 
a single, $T$-independent parameter governing the transformation.

\begin{figure*}[t]
\center
\includegraphics[width=0.32\linewidth]{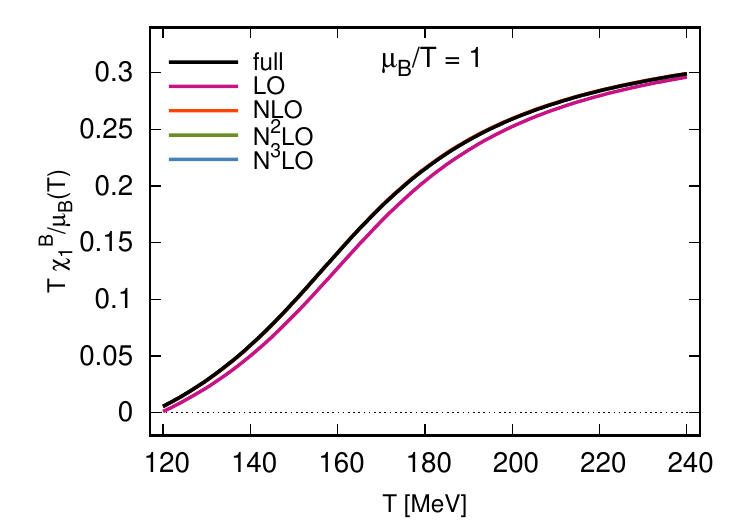}
\includegraphics[width=0.32\linewidth]{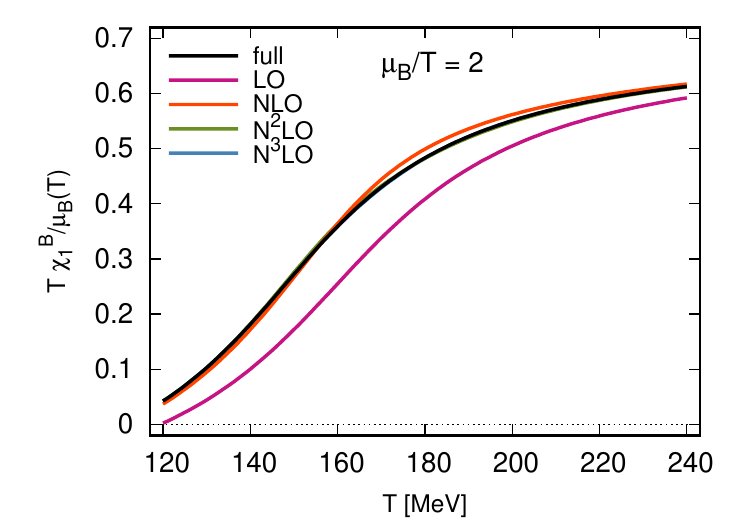}
\includegraphics[width=0.32\linewidth]{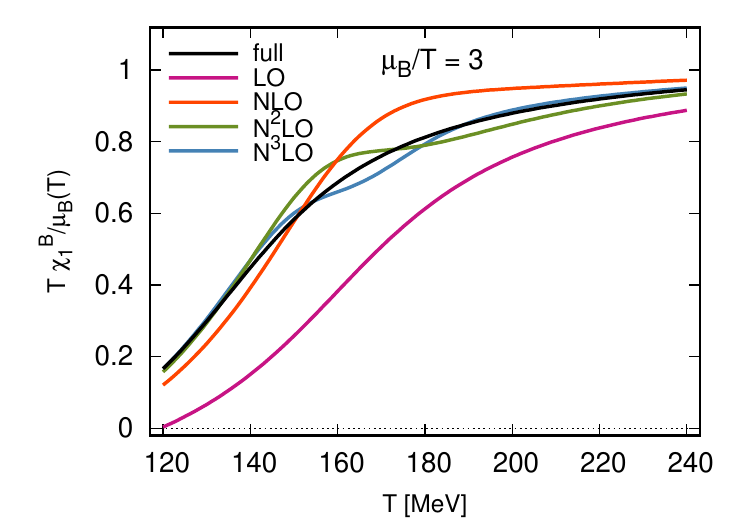}
\caption{Benchmarking various orders of the Taylor method assuming
an equation of state, where the $\mu_B$-dependence of $\chi^B_1/\hmu_B$ 
consists of a simple shift in temperature. This equation of state
is a somewhat simplified form of the observed behaviour.}
\label{fig:mock}
\end{figure*}

A similar behavior is observed for other quantities too. 
We show in Fig.~\ref{fig:shift_BS_S2} the first and 
second order fluctuations of strangeness at imaginary 
baryon chemical potentials. In analogy with 
Eq.~\eqref{eq:shift1} one has:
\begin{align}
\frac{\chi_1^S(T,\hmu_B)}{\hmu_B} &= \chi_{11}^{BS} (T^\prime,0) \, \,  , \\ \nonumber
\chi_2^S(T,\hmu_B) &= \chi_2^S (T^\prime,0) \, \,  , 
\end{align}
where $T^\prime$ is defined analogously to
Eq.~\eqref{eq:Tprime1}, albeit with different $\kappa$
parameters.

In order to motivate our alternative summation 
scheme, let us first consider a crude approximation
that we will later refine.
We take Eq.~(\ref{eq:Tprime1}) at face value and use it
together with Eq.~(\ref{eq:shift1}) to obtain a well defined
$\chi^B_1(T,\hmu_B)$ function. We need a $\chi^B_2(T,0)$
function as well, which we borrow from a deliberately
simple fit $f(T) = a + b \arctan(c(T-d))$ to our data
on a $\lat{48}{12}$ lattice. In principle, we could 
not only calculate $\chi^B_1(T,\hmu_B)$ at arbitrary 
$\hmu_B$ but, blindly believing Eq.~(\ref{eq:Tprime1}),
one could calculate the higher $\mu_B$-derivatives as well.
While this will not describe Nature precisely, it can
serve as a test for the Taylor expansion method.
To this end, we took several $\hmu_B$-derivatives of our
$\chi^B_1(T,\hmu_B)$ function and calculated its
truncated Taylor series for the lowest four orders.
Here LO means just plotting $\chi^B_2(T,0)$.
We compared our mock curve (labelled as ``full'') against 
its Taylor expansion at three real values of the chemical
potential (see Fig.~\ref{fig:mock}).

For $\hmu_B=1,2$, the summation up to LO and NLO
is sufficient to perfectly reproduce the function. 
However, as the chemical potential is increased, the 
Taylor expansion carried as far as the NNNLO does not 
reproduce the original function. On the one hand, 
convergence is achieved more slowly; on the other
hand, spurious effects appear, which generally manifest themselves
in a non-monotonicity of the function. These spurious effects in truncated Taylor series were pointed out also in Refs. \cite{Ratti:2007jf,Critelli:2017oub,Parotto:2018pwx}.

The picture emerging from this simple analysis is 
rather suggestive, especially when compared to the 
results shown in Fig.~\ref{fig:chiB1Taylor} (right panel)
obtained from actual lattice data. We also note that
this simple analysis does not suffer from the additional
complications of signal extraction for higher order 
expansion coefficients, which in turn play a relevant 
role in the real-data analysis.

\section{Formalism \label{sec:formalism}}

At vanishing chemical potential, it is possible to express the 
normalized baryon density as a Taylor expansion:
\beq  \label{eq:defin_1}
\frac{\chi_1^B}{\hmu_B} (T,\hmu_B) = \chi_2^B 
(T, 0) + \frac{\hmu_B^2}{6} \chi_4^B(T, 0) + \frac{\hmu_B^4}{120} \chi_6^B (T,0) + \cdots
\eeq 

As we saw in Fig.~\ref{fig:shift_B2}, the behavior 
of $\frac{\chi_1^B}{\hmu_B} (T,\hmu_B)$ at finite 
chemical potential clearly resembles that of 
$\chi_2^B (T, \hmu_B)$, although shifted/rescaled in 
temperature. As long as $\chi^B_1/\hmu_B$ is a monotonic function 
of $T$, the finite density physics can be encoded into the $T'(T,\hmu_B)$
function. A straightforward, but systematic generalization of 
Eq.~(\ref{eq:Tprime1}) reads:
\begin{equation} \label{eq:Tprime}
T^\prime (T,\hmu_B) = T \left( 1 + \kappa_2^{BB} (T) \hmu_B^2
+ \kappa_4^{BB} (T) \hmu_B^4 +\mathcal{O}(\hmu_B^6)\right) \, \, .
\end{equation}
In the above equation, we introduced the new parameters $\kappa_2^{BB}(T)$ and 
$\kappa_4^{BB}(T)$, which describe the shift/rescaling of the temperature of 
$\chi_1^B/\hmu_B$ at finite $\mu_B$. Analogous parameters will be introduced below 
to for the case of $\chi_1^S/\hmu_B$ ($\kappa_2^{BS}$ and $\kappa_4^{BS}$) and of 
$\chi_2^S$ ($\kappa_2^{SS}$ and $\kappa_4^{SS}$) at finite $\mu_B$.
In a way, this formalism replaces the fixed temperature $\mu_B$ expansion 
by a fixed-observable temperature expansion.

Having now two expressions, Eqs.~\eqref{eq:defin_1} and \eqref{eq:shift1},
for the same quantity we require their equality at each order in the
$\hmu_B$ expansion at $\mu_B=0$, having:
\begin{align} \label{eq:chisVkappas}
\chi_4^B (T) &= 6 T \kappa_2^{BB} (T) \frac{d\chi_2}{dT}  \, \, ,\\ \nonumber
\chi_6^B (T) &=  60 T^2 (\kappa_2^{BB})^2 (T) \frac{d^2\chi_2}{dT^2} + 120 T \kappa_4^{BB} (T) \frac{d\chi_2}{dT}  \, \, ,
\end{align} 
which in turn yields:
\begin{align} \label{eq:kappasVchis}
\kappa_2^{BB} (T) &= \frac{1}{6T} \frac{\chi_4^B(T)}{{\chi_2^B}^\prime(T)} \, \, , \\ \nonumber
\kappa_4^{BB} (T) &= \frac{1}{360 {{\chi_2^B}^\prime(T)}^3} \left( 3 {{\chi_2^B}^\prime(T)}^2 \chi_6^B(T) - 5 {\chi_2^B}^{\prime \prime}(T) {\chi_4^B(T)}^2 \right) \, \, .
\end{align}

A similar treatment can be applied to the other observables. 
For the second order fluctuations including baryon number and 
strangeness, one can consider:
\beq  \label{eq:defin_BS}
\frac{\chi_1^S}{\hmu_B} (T,\hmu_B) = \chi_{11}^{BS} 
(T, 0) + \frac{\hmu_B^2}{6} \chi_{31}^{BS}(T, 0) + \frac{\hmu_B^4}{120} \chi_{51}^{BS} (T,0) + \cdots
\eeq 
and:
\beq  \label{eq:defin_SS}
\chi_2^S (T,\hmu_B) = \chi_2^S (T, 0) + \frac{\hmu_B^2}{2} 
\chi_{22}^{BS}(T, 0) + \frac{\hmu_B^4}{24} 
\chi_{42}^{BS} (T,0) + \cdots
\eeq

Similarly as before, one can show that:
\begin{align}
\kappa_2^{BS} (T) &= \frac{1}{6T} 
\frac{\chi_{31}^{BS}(T)}{{\chi_{11}^{BS}}^\prime(T)}  \, \, ,
\\ \nonumber
\kappa_4^{BS} (T) &= 
\frac{1}{360{{\chi_{11}^{BS}}^\prime(T)}^3}  
\left( 3 {{\chi_{11}^{BS}}^\prime(T)}^2 \chi_{51}^{BS}(T) 
\right. \\  \nonumber
& \left. \qquad \qquad \qquad \qquad - 5 {\chi_{11}^{BS}}^{\prime \prime}(T) 
{\chi_{31}^{BS}(T)}^2 \right) \, \, ,
\end{align}
and: 
\begin{align}
\kappa_2^{SS} (T) &= \frac{1}{2T} 
\frac{\chi_{22}^{BS}(T)}{{\chi_2^S}^\prime(T)} \, \, , 
\\ \nonumber
\kappa_4^{SS} (T) &= 
\frac{1}{24{{\chi_2^S}^\prime(T)}^3}  
\left( {{\chi_2^S}^\prime(T)}^2 \chi_{42}^{BS}(T) 
\right. \\  \nonumber
& \left. \qquad \qquad \qquad \qquad - 3 {\chi_2^S}^{\prime \prime}(T) 
{\chi_{22}^{BS}(T)}^2 \right) \, \, .
\end{align}

\begin{figure}
\center
\includegraphics[width=\linewidth]{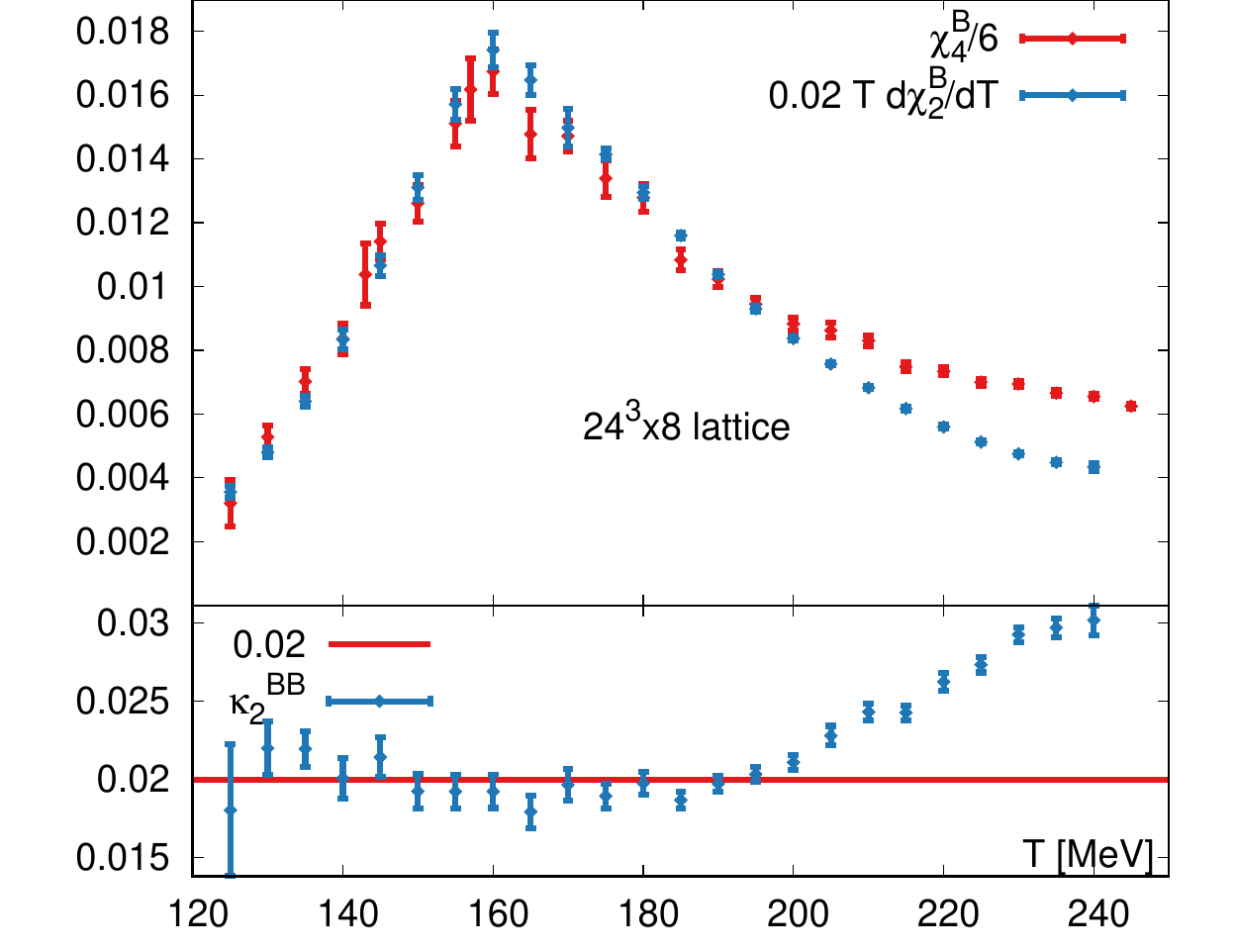}
\caption{\label{fig:chider_24x8}
Evaluation of Eq.~\eqref{eq:kappasVchis} on a coarse lattice with high
statistics. The resulting $\kappa_2^{BB}(T)$ shows a very mild temperature
dependence in the transition region.
}
\end{figure}

Before we embark into the discussion of the lattice analysis, let us
have an impression on the discussed quantities.
Eq.~\eqref{eq:kappasVchis} gives a way to directly 
determine $\kappa_2^{BB}(T)$ and $\kappa_4^{BB}(T)$ using only $\mu_B=0$ data. 

This approach might be subject to numerical problems, especially in
the case of $\kappa_4^{BB}(T)$, which is obtained as the difference of two competing
terms. Notice, too, that Eq.~\eqref{eq:kappasVchis} contains temperature-derivatives 
of the $\chi(T)$ coefficients, which may pose a numerical
challenge, unless the coefficients are known at sufficient statistics and
resolution in $T$.

On lattices where high statistics data taking is feasible, we can investigate
Eq.~\eqref{eq:kappasVchis}, at least for $\kappa_2^{BB}$. In the top panel of
Fig.~\ref{fig:chider_24x8} we compare the numerator and (rescaled)
denominator of Eq.~\eqref{eq:kappasVchis}, while their ratio $\kappa_2^{BB}(T)$ 
is shown in the bottom panel. In the entire transition region
the ratio is consistent with a constant, because the peak in $\chi^B_4(T)$ is
replicated in the temperature dependence of $\chi^B_2(T)$. As opposed to
the Taylor coefficients, $\kappa_2^{BB}(T)$ shows a very mild temperature dependence.

Finally we remark that very similar equations have been already used 
in Ref.~\cite{Bazavov:2017dus} to calculate ``lines of constant physics''
to $\mathcal{O}(\mu_B^4)$ order. In this reference the
pressure, energy density and entropy were calculated using the Taylor
method, and in a further step lines were drawn on the $\mu_B-T$ 
phase diagram, where these quantities are constant in some normalization.
The obtained $\kappa_2$ coefficients are closely related to ours. 

Contrary to Ref.~\cite{Bazavov:2017dus}, we use Eq.~(\ref{eq:Tprime}) as the
definition of a truncation scheme rather than to investigate a Taylor expanded
result. In a way, we work in the opposite direction: we will use lattice
data at zero and imaginary $\mu_B$ to obtain the coefficients in 
Eq.~(\ref{eq:Tprime}), which then can be used to either calculate the 
Taylor coefficients or, even better, to extrapolate the equation of state
at finite $\mu_B$ with no reference to the Taylor coefficients themselves.

Also, we used imaginary chemical potentials not only to calculate the
coefficients, but also to study the single observable first, on
which the analysis is based. 
We base our description of the entire chemical potential-dependence of 
the QCD free energy function on $\chi_1^B(T,\hmu_B)/\hmu_B$. It is essential
to rely on one observable only, in order to guarantee thermodynamic consistency: 
entropy, pressure and energy density will obey the known thermodynamic relations 
(see Section~\ref{sec:thermo}) only if they come from the same truncation scheme.

\section{Lattice determination of the expansion coefficients \label{sec:lattice}}

\subsection{Lattice details}

In this work, we use the lattice action and the parameters described in
Ref.~\cite{Bellwied:2015lba}. The action benefits from tree-level
Symanzik improvement in the gauge sector and four levels of stout smearing
for the staggered flavors. The up and down quarks are degenerate. The
resulting light pair of quarks, as well as the strange and charm quarks assume
their respective physical mass.

We performed simulations at $\mu_B=0$ on $32^3\times 8$, $40^3\times 10$, 
$48^3\times12$ and $64^3\times 16$ lattices in a temperature range of 
$130-300~\MeV$, and up to $500 \MeV$ on larger volumes. These simulations 
were complemented
at imaginary values of the chemical potential in the temperature range 
$135-245~\mathrm{MeV}$ for the lattice resolutions $N_\tau=8,\dots,12$.
The $\mu_B\ne0$ data were simulated at $\mu_S=0$, some of these ensembles were
already used in Ref.~\cite{Borsanyi:2018grb}.

In addition, we performed a high-statistics run on the cheap and coarse
$24^3\times 8$ lattice, mainly to produce Fig.~\ref{fig:chider_24x8}.
These data did not enter the continuum extrapolation.

\subsection{The coefficients $\kappa_2^{ij}$ and $\kappa_4^{ij}$}

\begin{figure}[!]
\includegraphics[width=\linewidth]{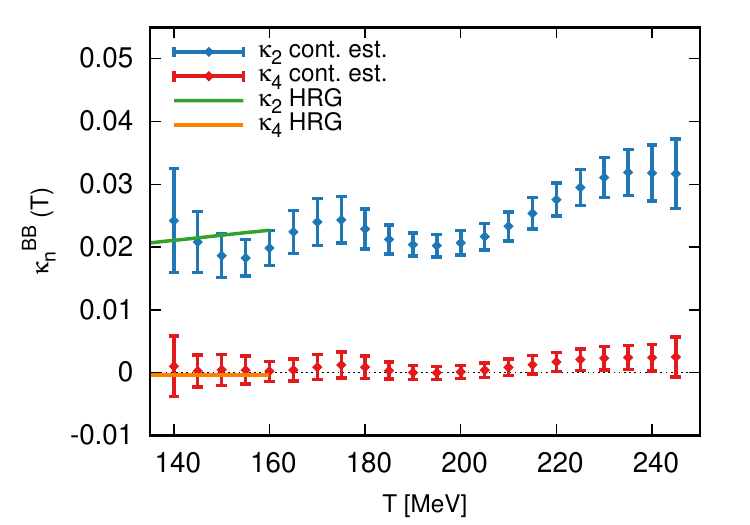}
\includegraphics[width=\linewidth]{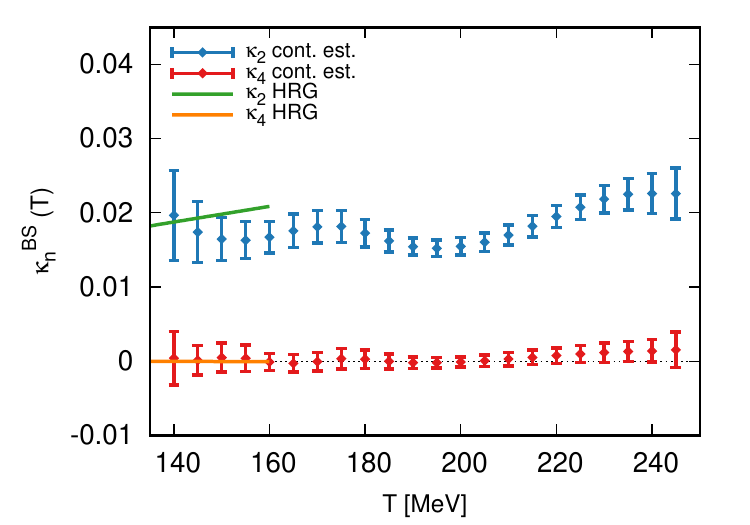}
\includegraphics[width=\linewidth]{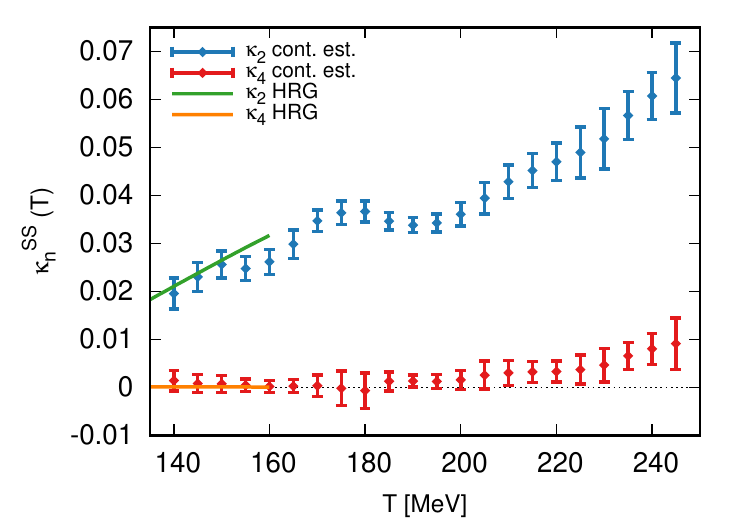}
\caption{Continuum extrapolated result for the parameters $\kappa_2^{BB} (T)$
and $\kappa_4^{BB} (T)$ (top panel), $\kappa_2^{BS} (T)$ and $\kappa_4^{BS}
(T)$ (central panel), and $\kappa_2^{SS} (T)$ and $\kappa_4^{SS} (T)$ (bottom
panel). The parameters $\kappa_2^{ij}$ are shown in blue, and the
$\kappa_4^{ij}$ in red.  HRG results for all quantities are shown up to
$T=160\MeV$ (in green for $\kappa_2^{ij}$ and orange for $\kappa_4^{ij}$,
respectively).} 
\label{fig:final_kN}
\end{figure}

For the determination of $\kappa_2^{ij}$ and $\kappa_4^{ij}$, 
one can take advantage of simulations both at zero and 
finite imaginary chemical potential. Using 
Eq.~\eqref{eq:kappasVchis} we calculated $\kappa_2^{BB}(T)$, shown
in Fig.~\ref{fig:chider_24x8}. To extract $\kappa_4^{BB}(T)$ using the same strategy,
a precise result on $\chi^6_B(T)$ would be necessary.

Instead, we utilize imaginary chemical potential simulations as 
follows. We perform simulations at imaginary values of 
the baryon chemical potential:
\begin{equation}
\hmu_B = i \frac{n\pi}{8} \, \, , \qquad n \in \{3,4,5,6\}
\end{equation}
with $\hmu_Q = \hmu_S = 0$.

We simulate temperatures in the range 
$T = 135 - 245 \MeV$. For each temperature $T$ and 
chemical potential $\hmu_B$ we determine the 
temperature $T^\prime$ for which Eq.~\eqref{eq:defin_1}
holds, hence defining a function $T^\prime (T,\hmu_B)$. 
Rearranging the terms in Eq~\eqref{eq:Tprime}, and in a similar way for the other observables, one can write:
\begin{eqnarray}
\Pi(T,\hmu_B)&=&
\kappa_2^{ij}(T^\prime) + \kappa_4^{ij}(T^\prime) \hat{\mu}^2_B + \kappa_6^{ij}(T^\prime)\hat{\mu}^4_B +\dots \qquad 
\end{eqnarray}
with the proxy quantity
\begin{eqnarray}
\Pi(T,\hmu_B)&=&
\frac{T^\prime(T,\hmu_B) - T}{T^\prime(T,\hmu_B) \hat{\mu}_B^2}. 
\label{eq:proxy}
\end{eqnarray}
The proxy $\Pi(T,\hmu_B)$ can be determined
using lattice simulations. The relatively precisely
known function $\chi^B_2(T,0)$ is first
interpolated in temperature. Afterwards, we only need to measure 
$\chi_1^B(T,\mu_B)/\hmu_B$
on an ensemble with imaginary $\mu_B$.
Equating this to $\chi^B_2(T^\prime,0)$ gives us 
$T^\prime(T,\mu_B)$, while we have to take care of the propagation
of the statistical errors.

Having determined $\Pi(T,\hmu_B)$ for several imaginary chemical
potentials and several lattice spacings for each given temperature,
one can perform a polynomial fit in $\hmu_B^2$ and obtain the
expansion coefficients.  This can be done separately for each lattice. 
However, we prefer to combine the $\hmu_B$ and continuum fits in one
two-dimensional fitting procedure. This combined fit is repeated for
every temperature, in steps of 5~MeV.

%


In order to estimate the systematic uncertainties 
associated to our results, we perform a number
of analyses at each temperature.
There are several ambiguous points that need to be considered.
Most obviously, one could choose to include the $\kappa_6^{ij}(T)$
term in the fit or not,
or consider the fit of $1/\Pi$ instead.
Also the range in imaginary $\mu_B$ is arbitrary: we
consider $\mathrm{Im}~\mu_B\le 2.0$ or $\mathrm{Im}~\mu_B\le 2.4$.
When different lattice spacings are fitted together in a continuum
extrapolation, one selects the bare parameters such that the ensembles
correspond to the same physical temperature. This choice is, however,
ambiguous too, as the scale setting may be based on various observables. In
our case, we consider $f_\pi$ or $w_0$ to this purpose. As we mentioned before, 
$\chi^B_2(T,0)$ is subject to an interpolation, performed through basis splines. The 
same is true for $\chi^B_1(T,\hmu_B)$ at finite chemical potentials. Since the location 
of the node points is also arbitrary, we include three versions at $\mu_B=0$ and two 
at imaginary $\mu_B$, each with perfect fit quality. Finally, in the continuum 
extrapolation the coarsest lattice, $32^3\times8$, has either been used or dropped.
The listed options can be considered in arbitrary combinations. In total
we carry out all 144 fits to perform a continuum extrapolation of $\kappa_2^{ij}(T)$ 
and $\kappa_4^{ij}(T)$. After dropping the fits with a Q-value below a percent,
we use uniform weights to produce histogram out of these (somewhat less than)
144 results for each temperature. The width of the histogram defines
the systematic error. In the plots we show combined errors, where we assume
that statistical and systematic errors add up in quadrature.
This systematic error estimation procedure has been used and described
in more detail in several of our works, most recently in
Ref.~\cite{Borsanyi:2018grb}.

In the top panel of Fig.~\ref{fig:final_kN} we show the results
of the temperature-by-temperature fit procedure for the 
parameters $\kappa_2^{BB} (T)$ and $\kappa_4^{BB} (T)$, alongside
the corresponding expectations from the Hadron Resonance Gas (HRG) model. 
We find that, within errors, $\kappa_2^{BB}(T)$ has hardly any dependence on
the temperature, while $\kappa_4^{BB} (T)$ is everywhere consistent with zero
at our current level of precision. Nonetheless, a clear separation of almost
one order of magnitude appears between these two coefficients. We also note
that good agreement with the HRG results is found up to at least $T=160\MeV$.

In the central and bottom panels of Fig.~\ref{fig:final_kN} we show our 
results for the parameters $\kappa_2^{BS}$, $\kappa_4^{BS}$ and
$\kappa_2^{SS}$, $\kappa_4^{SS}$ respectively, together with their HRG
determinations. While for $\chi_{11}^{BS}$ barely any temperature dependence is
observed, as in the case of $\chi_2^B$, for $\chi_2^S$ a much stronger
$T$-dependence is clearly visible. As in the case of $\kappa_4^{BB}$,
$\kappa_4^{BS}$ is consistent with zero throughout the whole temperature range
we consider. On the other hand, $\kappa_4^{SS}$ rises above zero for
temperatures $T \gtrsim 220 \MeV$.

\begin{figure}[!]
\includegraphics[width=\linewidth]{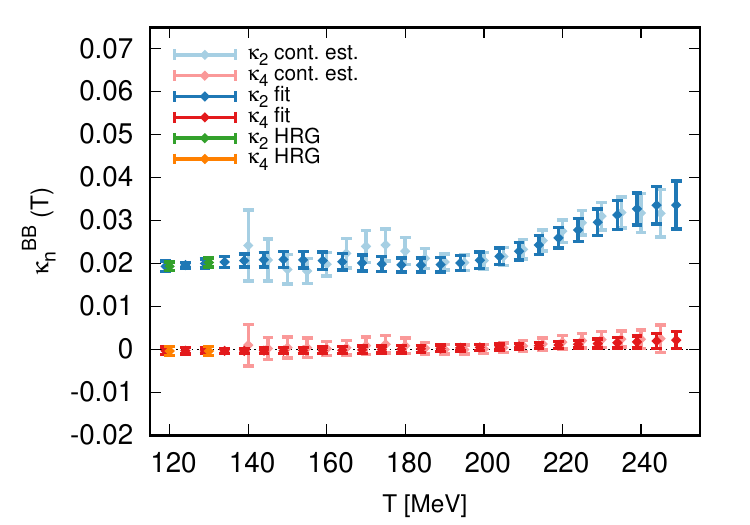}
\includegraphics[width=\linewidth]{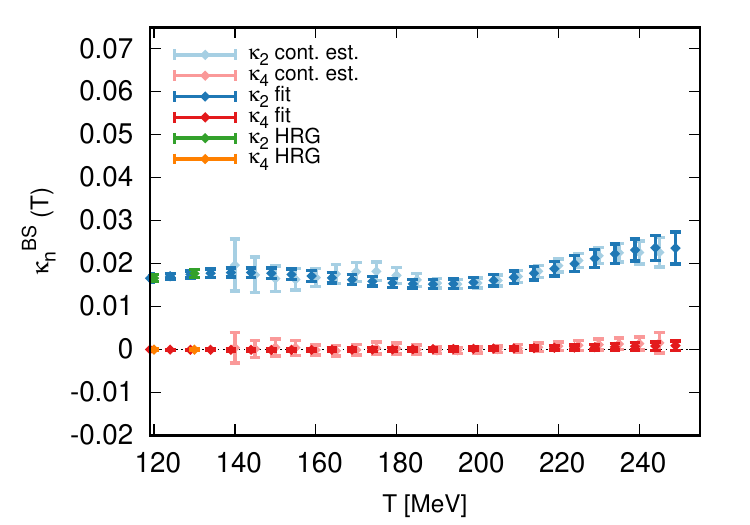}
\includegraphics[width=\linewidth]{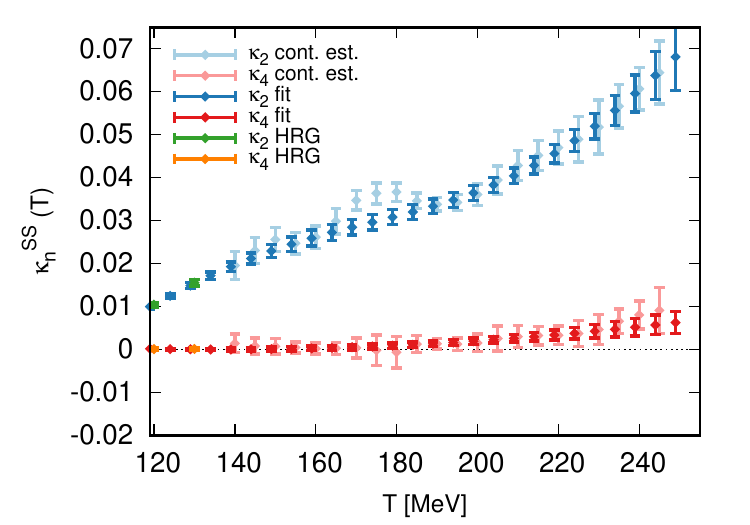}
\caption{Results of the polynomial fits to $\kappa_2^{BB}$ and 
$\kappa_4^{BB}$ (top panel),  $\kappa_2^{BS} (T)$ and $\kappa_4^{BS} (T)$
(central panel), and $\kappa_2^{SS} (T)$ and $\kappa_4^{SS} (T)$ (bottom
panel). The parameters $\kappa_2^{ij}$ are shown in blue, and the
$\kappa_4^{ij}$ in red. The fitted quantities from Fig.~\ref{fig:final_kN}
are shown by lighter blue and red 
points. Due to the lack of points at low $T$, we continued the data
set with HRG points, show as green and orange dots.
Though the polynomials do not always follow all ``excursions'' of the
$T$-by-$T$ result, the reduced $\chi^2$ of the correlated temperature
fit is always $<1$. This is possible, because the fitted data are highly
correlated, coming from both statistical and systematic effects.}
\label{fig:fits_kN}
\end{figure}

%
%

Note that the error bars in these plots are highly correlated. The correlation
is mostly systematic. The apparent `waves' are often statistically not 
as significant as it seems at a first glance. For comparison we refer to the
direct result on our coarsest lattice in Fig.~\ref{fig:chider_24x8},
where these `waves' are absent.
Before moving on to calculate thermodynamic observables at finite real
chemical potential, we construct a smoother version of our final results for
$\kappa_2^{ij}$ and $\kappa_4^{ij}$, in order to limit the influence of
numerical effects on the final observables, but also to facilitate the
temperature derivatives that enter the entropy formula.
Thanks to the very mild dependence of the coefficients 
on the temperature, we perform a polynomial fit of order 5 for the
$\kappa_2^{ij}$, and of order 2 for the even less $T$-dependent
$\kappa_4^{ij}$. The very good fit qualities did not motivate higher
order polynomials, using fourth or sixth order for $\kappa_2^{ij}$ hardly
changes the result.  In order to stabilize the low-temperature behavior, we
include in the fit two points from the HRG model for $T = 120, 130 \MeV$, to
which we associate an uncertainty of $5\%$ for $\kappa_2^{ij}$ and of $300\%$
for $\kappa_4^{ij}$.  The choice of these particular values for the
uncertainties is uniquely guided by the necessity of placing a constraint on
the low-T behavior, while avoiding to drive the fit too strongly. For this
reason, these arbitrary errors are chosen to be smaller, but comparable to the
lattice ones. We note that the fits we perform take fully into account the
correlations between results at different temperatures, systematic as well as
statistical. Thus we encode all errors into the (correlated) errors of the
coefficients of a polynomial.

In Fig.~\ref{fig:fits_kN} we show the results of the fits in darker color, with 
the fitted data in lighter shades. The HRG points included in the fit are shown as well.

\subsection{Continuum result on $\chi^B_2(T)$ and its temperature derivative}

\begin{figure}
\begin{center}
\includegraphics[width=\linewidth]{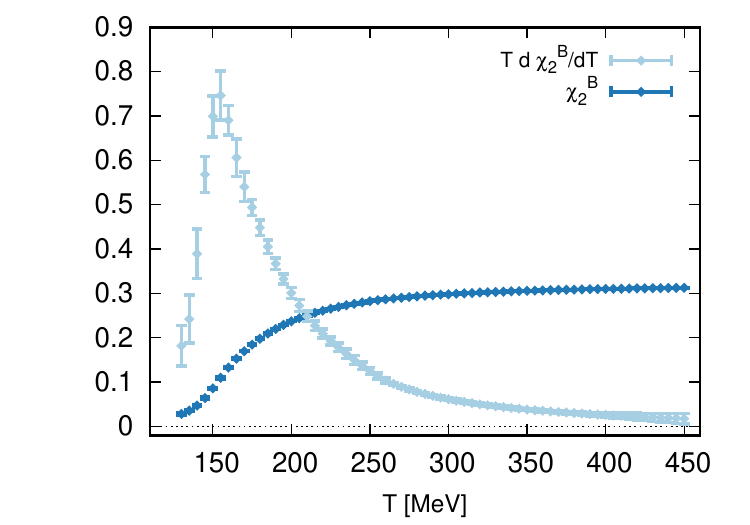}
\caption{\label{fig:chider}
Continuum extrapolated $\chi^B_2(T)$ and $T d\chi^B_2(T)/dT$ functions
from $N_\tau=10$, 12 and 16 lattices at $\mu_B=0$. Together with 
$\kappa^{BB}_n(T)$ these functions enter the thermodynamic analysis.
}
\end{center}
\end{figure}

In order to determine the value of thermodynamic quantities at finite real
chemical potential, a continuum result for the basic quantity $\chi^B_2(T,0)$
is required in addition to $\kappa_2^{BB}(T)$ and $\kappa_4^{BB}(T)$.  For some
observables like the entropy, the temperature derivative of $\chi^B_2(T)$ 
has to be calculated
as well.  We have already published this quantity in
Ref.~\cite{Bellwied:2015lba}, but
not its temperature derivatives.  To make this work self-contained, we update
this determination with the updated statistics using lattices up to
$64^3\times16$.

We have divided the temperature range into two parts: the transition region and
the higher, near-perturbative temperatures. For the lower temperature part,
we interpolate using basis splines using the 
$T\in [130~\mathrm{MeV},300~\mathrm{MeV}]$ data points. For the quantity 
$T d\chi^B_2(T)/dT$ we extended the data range with the HRG curve, so that the
numerical derivative has a level arm at the lowest temperatures. 
The basis splines are cubic splines in
the given temperature range.  We fit the splines in temperature and in
$1/N_\tau^2$ in one step. The fitted function in this range is thus:
\begin{equation}
\chi^B_2(T,0;N_\tau) = \sum_{i=1}^n \alpha_i b_i(T) + 
\frac{1}{N_\tau^2}\sum_{i=1}^n \beta_i b_i(T) \,,
\label{eq:splinecont}
\end{equation}
where $b_i(t_j) = \delta_{ij}$ and $t_j$ are the knots for the spline with $j=1\dots n$.
$n$ and the $t_j$ set are the only arbitrary inputs. We combine the result from
four sets of values for them, so that we can estimate the systematics. Only the lattices with 
$N_\tau=10,12$ and 16 enter the continuum extrapolation. For 
$\chi^B_2(T,0;N_\tau)$ in Eq.~\eqref{eq:splinecont},
the tree-level improvement was already applied \cite{Bellwied:2015lba}.
The lattice resolutions and the spline knots are selected such that
the fit qualities (Q value) are distributed as expected.  We also took into
account the scale setting ambiguity by combining the results with $w_0-$ and 
$f_\pi-$based scale settings. The systematics of the $T$-derivative 
$T d \chi^B_2(T,0)/dT$ and of $\chi^B_2(T,0)$ were determined separately.

In the high temperature regime, the smooth monotonic behavior is not well
described with cubic splines. Instead we performed a high order polynomial fit
in the inverse temperature $1/T$. The known analytical form of the convergence
to the Stefan-Boltzmann limit is, of course, not this polynomial. We do not
wish to enforce the perturbative behavior at the intermediate temperatures that
we describe. Thus, the constant in the $1/T$ description is not exactly the
Stefan-Boltzmann limit, and for this reason, we cannot regard this as a basis
for an extrapolation. However, this simple approach allows the interpolation
and the calculation of the derivative. We used lattice data in the range
$T=180-450~\mathrm{MeV}$.

High- and low-temperature regions overlap and the two fitting methods
give consistent results between $T=200 - 280 \MeV$ for both quantities.
Thus, we simply concatenate the resulting functions at $T=260 \MeV$. 
We show the final version of the $T d\chi^B_2(T,0)/dT$ and
$\chi^B_2(T,0)$ functions in Fig.~\ref{fig:chider}.

\section{Thermodynamics at real chemical potential\label{sec:thermo}}

\begin{figure*}[!]
\includegraphics[width=0.49\linewidth]{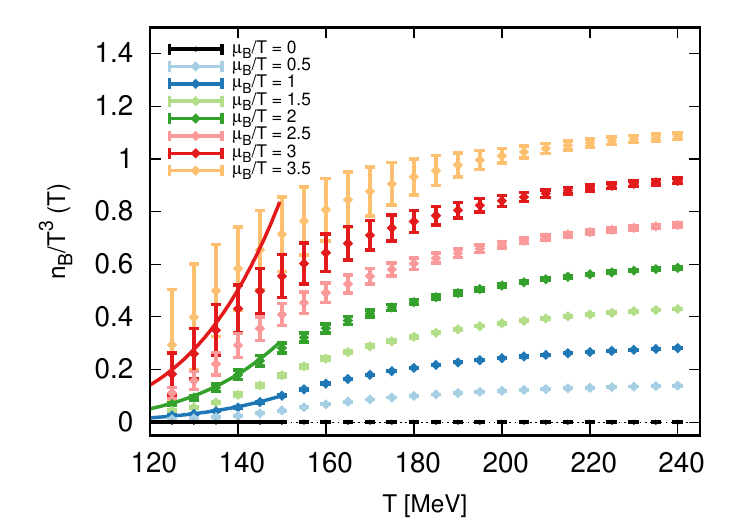}
\includegraphics[width=0.49\linewidth]{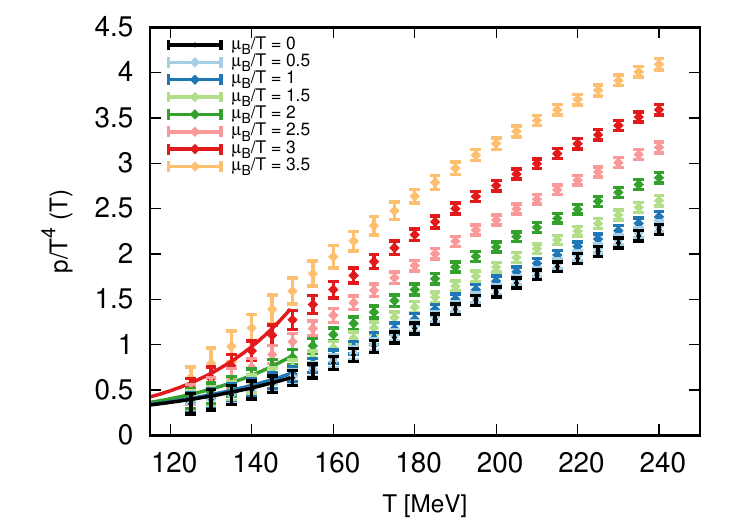}
\includegraphics[width=0.49\linewidth]{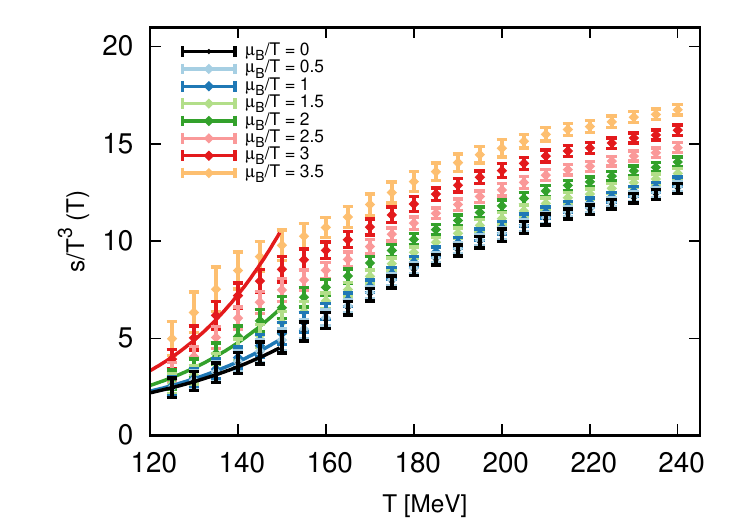}
\includegraphics[width=0.49\linewidth]{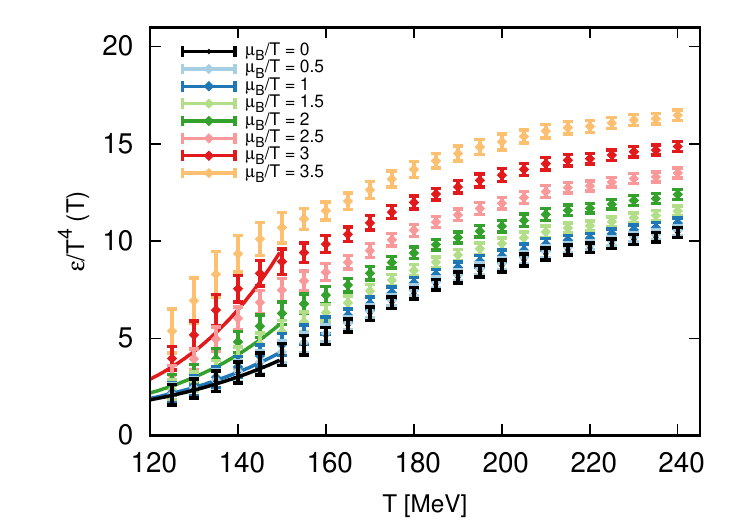}
\includegraphics[width=0.49\linewidth]{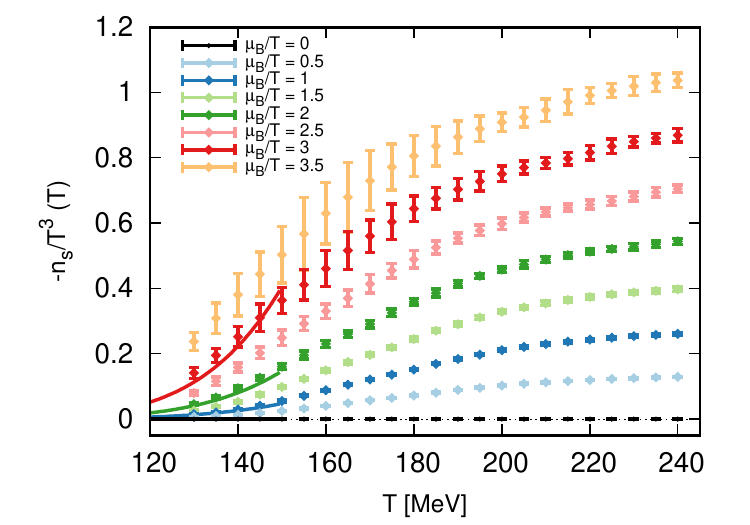}
\includegraphics[width=0.49\linewidth]{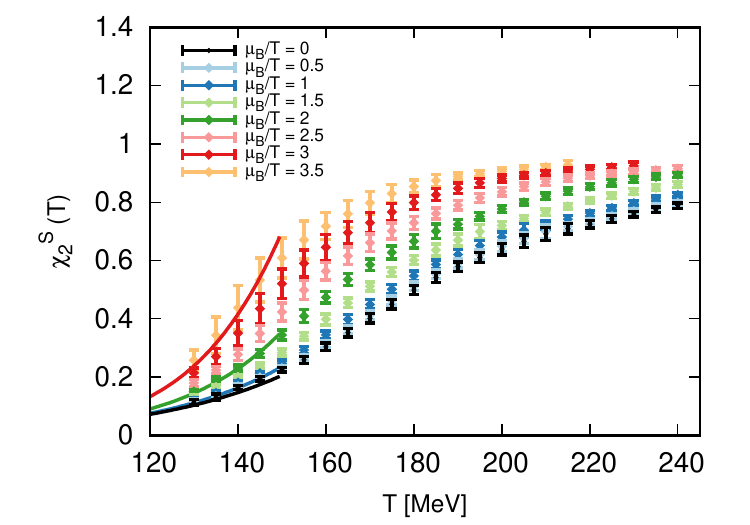}
\caption{Baryon density, pressure, entropy, energy density, strangeness density and 
$\chi_2^S$ at increasing values of $\hmu_B$. With solid lines we show the results from
the HRG model.}
\label{fig:thermo_mux}
\end{figure*}

Once $n_B$ is determined, we have everything we need to extract the other 
thermodynamic quantities. The integration constant for the pressure is obviously the 
pressure itself at $\mu_B=0$.

We note here that, on the lattice, we always deal with 
dimensionless thermodynamic quantities, which 
correspond to the physical ones divided 
by suitable powers of the temperature. E.g., the 
dimensionful baryon density is 
$n_B = T^3 \hat{n}_B$ (we will hereafter use the hat to 
indicate dimensionless quantities).

From the baryon density $\hat{n}_B(\hmu_B,T)$, the pressure is 
obtained through simple integration:
\begin{equation}
\frac{p(\mu_B,T)}{T^4} = \hat{p}(\hmu_B,T) = \hat{p}(0,T) + \bigintssss_0^{\hmu_B} \!\!\!\!\!\! {\rm d} \hmu^\prime_B \, \hat{n}_B (\hmu_B^\prime,T) \, \, .
\end{equation}

The entropy density is defined as: 
\begin{equation}
s (\mu_B,T) = \left. \frac{\partial p(\mu_B,T)}{\partial T} \right|_\mu. 
\end{equation}
For dimensionless quantities:
\begin{align}
\hat{s}(\hmu_B,T) &= 4 \, \hat{p}(\hmu_B,T) + T  \left. \frac{\partial \hat{p}(\hmu_B,T)}{\partial T} \right|_{\mu} = \\ \nonumber
&= 4 \, \hat{p}(\hmu_B,T) + T  \left. \frac{\partial \hat{p}(\hmu_B,T)}{\partial T} \right|_{\hmu} - \hmu_B \hat{n}_B(\hmu_B,T)
\end{align}
where in the last step we converted the derivative 
at constant $\mu_B$ into a derivative at constant 
$\hmu_B$.

The $T$-derivative of the pressure involves the chain rule
\begin{align}
& \quad T  \left. \frac{\partial \hat{p}(\hmu_B,T)}{\partial T} \right|_{\hmu} = T  \left. \frac{\partial \hat{p}(0,T)}{\partial T} \right|_{\hmu}  \\
& \quad \qquad + \frac12 \int_0^{\hmu_B^2} T \left.\frac{d \chi^B_2(T^\prime)}{dT^\prime}\right|_{T^\prime = T\left( 1+\kappa_2^{BB} y + \kappa_4^{BB}y^2 \right)} \times \nonumber \\
& \quad \times \left[ 1+\kappa_2^{BB}y +\kappa_4^{BB}y^2 + T \left( \frac{d\kappa_2^{BB}}{dT} y+\frac{d\kappa_4^{BB}}{dT}y^2 \right) \right] dy \nonumber
\end{align}
where $\frac{d \chi^B_2(T)}{dT}$ is calculated at $\mu_B=0$ as already described.

The dimensionless expression for the energy density $\hat \epsilon= \epsilon/T^4$ follows as:
\begin{equation}
\hat{\epsilon}(\hmu_B,T) = \hat{s}(\hmu_B,T) - \hat{p}(\hmu_B,T) + 
											\hmu_B \hat{n}_B(\hmu_B,T).
\end{equation}


%

\begin{figure*}
\includegraphics[width=0.48\linewidth]{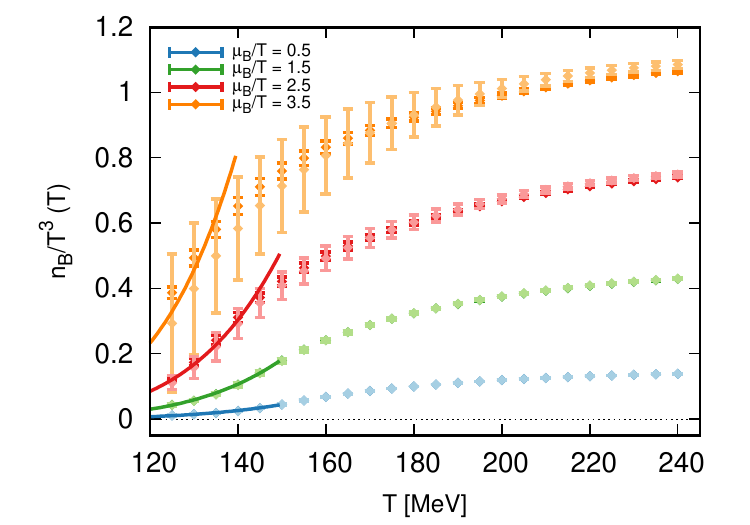}
\includegraphics[width=0.48\linewidth]{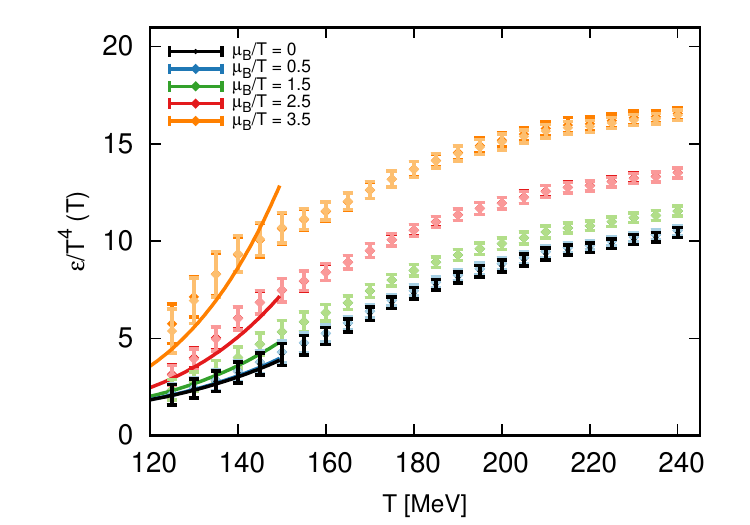}
\caption{Comparison of baryon density (left) and energy density (right) at different 
values of $\hmu_B$ in the case where a $\kappa_4^{BB}$ parameter is used (darker shades)
or omitted (lighter shades). The HRG results are shown with solid lines.}
\label{fig:bdens_mux}
\end{figure*}

Finally, we present our results for the finite real chemical potential
extrapolation of several thermodynamic quantities. The various panels of Fig. \ref{fig:thermo_mux} show 
the baryon density, pressure, entropy, energy density, strangeness density and 
$\chi_2^S$ for $\hmu_B = 0 - 3.5$.  Alongside our results, we show predictions
from the HRG model for $T < 150 \MeV$, which we find in very good agreement 
with our extrapolation for all observables, at all values of the chemical potential.

We also note that in all cases, the observables do not suffer from pathological 
behavior. The uncertainties are under control for our range of chemical potentials, 
which highly improves on the results currently achievable via Taylor expansion. 

We devote the two panels of Fig.~\ref{fig:bdens_mux} to the comparison of
our results for the baryon density (left) and energy density (right) to the simplified 
case where $\kappa_4^{BB}$ is neglected. We can appreciate how the inclusion of 
the next-to-leading-order parameter came at the cost of an increased uncertainty
at larger chemical potential.  This does not come unexpected, as we saw from our 
results that $\kappa_4^{BB}$ was compatible with zero at all temperatures. In the case 
of the energy density, which is dominated by the $\mu_B=0$ contribution, hardly any 
effect is visible.

%
%
%
%
%
%
%

\section{Conclusions\label{sec:conclusions}}

In this work, we proposed an alternative summation scheme for the equation
of state of QCD at finite real chemical potential, designed to overcome the 
shortcomings which are characteristic of the Taylor expansion approach. Combining
simulations at both zero and imaginary chemical potentials, we determined 
the LO and NLO parameters describing the chemical potential dependence of the 
baryon density, which we could then extrapolate to large real chemical potentials.

By combining this new element, and previously published results for the EoS at vanishing density, we 
could reconstruct all thermodynamic variables at chemical potentials as large as
$\mu_B/T = 3.5$ with rather limited uncertainty. Systematic as well as statistical errors
were considered in the analysis. 

These results, although still limited in precision at the level of the parameters 
$\kappa_2^{ij}$ and $\kappa_4^{ij}$, suggest that the avenue we pursue in this work is 
rather promising for the description of QCD thermodynamics at finite chemical 
potential. Moreover, our procedure is systematically improvable with sufficient 
computing power, and might prove to be a better strategy than existing ``canonical''
approaches.

In this work we limited ourselves to the case where the strange and electric
chemical potentials are set to zero. We reserve for future work the exploration
of the phenomenologically relevant case of strangeness neutrality and
fixed electric charge to baryon ratio.

\subsection*{Acknowledgments}
This project was funded by the DFG grant SFB/TR55.
The project also received support from the BMBF Grant No. 05P18PXFCA.
This work was also supported by
the Hungarian National Research,  Development and Innovation Office, NKFIH
grant KKP126769. A.P. is supported by the J. Bolyai Research
Scholarship of the Hungarian Academy of Sciences and by the \'UNKP-20-5 New
National Excellence Program of the Ministry for Innovation and Technology.
The project leading to this publication has received funding from
Excellence Initiative of
Aix-Marseille University - A*MIDEX, a French ``Investissements d'Avenir''
programme, AMX-18-ACE-005.
This  material is based upon  work  supported  by  the National  Science
Foundation under grants no. PHY-1654219 and by the U.S. DoE, 
Office  of  Science,  Office  of  Nuclear  Physics, within the framework of the
Beam Energy Scan Topical (BEST) Collaboration.  
This research used resources of the Oak Ridge Leadership Computing Facility, which is a DOE Office of Science User Facility supported under Contract DE-AC05-00OR22725. 
The authors gratefully
acknowledge the Gauss Centre for Supercomputing e.V.  (www.gauss-centre.eu) for
funding this project by providing computing time on the GCS Supercomputer
HAWK at HLRS, Stuttgart. 
Part of the computation was performed on the QPACE3 funded by the DFG ind
hosted by JSC.
C.R. also acknowledges the support from the Center of Advanced
Computing and Data Systems at the University of Houston.

\bibliography{reference}

\end{document}